\documentclass[reprint,groupedaddress,amsmath,amssymb,aps,showkeys]
{revtex4-1}

\usepackage{graphicx}
\usepackage{dcolumn}
\usepackage{bm}
\usepackage{hyperref}
\usepackage{physics}
\usepackage{mathtools}
\usepackage{subcaption}

\def\p{\partial}

\usepackage{graphicx,amssymb,amsmath,amsthm,amsfonts}
\usepackage[usenames]{color}
\usepackage{xcolor}
\usepackage{mathtools}

\begin{document}

\title{Quasinormal modes of Proca fields in a Schwarzschild-AdS
  spacetime}

\author{Tiago V. Fernandes}
\affiliation{Centro de Astrof\'{\i}sica e Gravita\c c\~ao  - CENTRA,
Departamento de F\'{\i}sica, Instituto Superior T\'ecnico - IST,
Universidade de Lisboa - UL,
Avenida Rovisco Pais 1, 1049-001 Lisboa, Portugal}
\author{David Hilditch}
\affiliation{Centro de Astrof\'{\i}sica e Gravita\c c\~ao  - CENTRA,
Departamento de F\'{\i}sica, Instituto Superior T\'ecnico - IST,
Universidade de Lisboa - UL,
Avenida Rovisco Pais 1, 1049-001 Lisboa, Portugal}
\author{Jos\'{e} P. S. Lemos}
\affiliation{Centro de Astrof\'{\i}sica e Gravita\c c\~ao  - CENTRA,
Departamento de F\'{\i}sica, Instituto Superior T\'ecnico - IST,
Universidade de Lisboa - UL,
Avenida Rovisco Pais 1, 1049-001 Lisboa, Portugal}
\author{V\'itor Cardoso}
\affiliation{Centro de Astrof\'{\i}sica e Gravita\c c\~ao  - CENTRA,
Departamento de F\'{\i}sica, Instituto Superior T\'ecnico - IST,
Universidade de Lisboa - UL,
Avenida Rovisco Pais 1, 1049-001 Lisboa, Portugal}

\begin{abstract}

We present new results concerning the Proca massive vector field in a
Schwarzschild-anti-de Sitter (Schwarzschild-AdS) black hole
geometry.  We provide a first principles
analysis of Proca vector fields in this geometry using both the
vector spherical harmonic (VSH) decomposition and separation method
and the Frolov-Krtou\v{s}-Kubiz\v{n}\'{a}k-Santos (FKKS) method that
separates the relevant equations in spinning geometries. The analysis
in the VSH method shows, on one hand, that it is arduous to separate
the scalar-type from the vector-type polarizations of the electric
sector of the Proca field, and on the other hand, it displays clearly
the electric and the magnetic mode sectors. The analysis in the FKKS
method is performed by taking the nonrotating limit of the Kerr-AdS
spacetime, and shows that the Ansatz 
decouples the polarizations in the electric mode sector even in
the nonrotating limit. On the other hand, it captures only two of
the three possible polarizations; indeed, the magnetic mode sector,
which is of vector-type, is missing. The reason for the
absence of the remaining polarization is related to the degeneracy of
the principal tensor in static spherical symmetric spacetimes.
The degrees of freedom and quasinormal modes in both
separation methods of the Proca field
are found.
The frequencies of the quasinormal modes are also carefully computed.
For the electric mode sector in the VSH method the frequencies are
found through an extension, which substitutes  number
coefficients by matrix coefficients, of the Horowitz-Hubeny numerical
procedure, whereas for the magnetic mode sector in the VSH method
and the electric sector of the FKKS method
it is shown that a direct use of the
procedure can be made. 
The values of the
quasinormal mode frequencies obtained for each  method are
compared and shown to be in good agreement with each other. This
further supports the analytical approaches presented here for the
behavior of the Proca field in a Schwarzschild-AdS black hole
background.

\centerline{}

\centerline{}

\end{abstract}


\maketitle

\section{Introduction}\label{sec:Intro}

Vector fields are known to describe the electroweak and the strong
interactions of the standard model of particle physics. In particular,
the mediator of electromagnetic interactions is the photon and it can
be identified as an excitation of an Abelian massless vector
field. Since the standard model does not explain the existence of dark
matter, new kinds of fields have been proposed as candidates. Such
fields would have very feeble interactions with ordinary matter, but
would interact gravitationally. Thus, bodies with strong gravitational
effects serve as devices to probe the existence of such new fields.
General relativistic black holes and their dynamical description are
thus important to understand the behavior of possible new matter, such
as massive vector fields that obey the Proca equations.

The analysis of the Maxwell massless vector field in static spacetimes
has been performed using the vector spherical harmonics method, or VSH
method for short. In Schwarzschild spacetimes, the Maxwell equations
were separated, decoupled, and reduced to a single master equation
in~\cite{Ruffini:1973}.  In
Schwarzschild-anti-de Sitter (Schwarzschild-AdS) spacetimes, the Maxwell
equations and their quasinormal mode content were studied using
reflective boundary conditions at infinity with VSH techniques
in~\cite{Cardoso:2001,ckl}, and using vanishing energy flux boundary
conditions at infinity, the quasinormal modes were found in
\cite{WangHerdeiroSampaio,ChenChoCornell}.  The massive Proca field in
Schwarzschild-type
spacetimes with an interest on its late time behavior
was studied in~\cite{KZM}, and a quasinormal
mode analysis for Schwarzschild 
spacetimes was performed in~\cite{Rosa:2012}. Quasinormal modes in
Schwarzschild-AdS spacetimes with a focus on their monopole term were
analyzed in~\cite{Konoplya:2006}.

The separation of massless vector fields, such as the Maxwell field,
in the Kerr spacetime used a completely different approach, that of
the Newman-Penrose formalism, despite the reduced group of explicit
symmetries~\cite{Teukolsky:1972}. This already hinted at the
existence of another kind of symmetry.  Massive vector fields were
considered not only under
a small rotation approximation with the
equations yielding a ladder of coupled multipoles~\cite{Pani:2012vp},
but also without approximations using a fully numerical
approach~\cite{Witek:2012tr,Cardoso:2018tly}.
In a different development, and following previous
work~\cite{Frolov:2008jr,Houri:2008ng,Frolov:2017whj,
Frolov:2017,Lunin:2017} that takes into account the presence of the
principal tensor in the Kerr-NUT-AdS and Kerr-NUT-dS spacetimes, i.e.,
spacetimes describing a rotating black hole in four
and  higher
dimensions and that include the NUT parameters and a cosmological
constant, Frolov, Krtou\v{s}, Kubiz\v{n}\'{a}k, and
Santos~\cite{Frolov:2018}, or FKKS for short, were able to extend the
perturbation analysis to the case of a Proca massive vector field in
spinning geometries.
It was further shown that with
this formalism the Proca equations can be separated.
It is still
unclear whether the FKKS method covers all the degrees of freedom,
including polarizations and quasinormal modes, of the Proca field in
the whole Kerr-NUT-AdS and Kerr-NUT-dS family.
However, with the help of an analysis for the 
marginally bound state case~\cite{Dolan:2018dqv,Baumann:2019eav},
it was found that for the Kerr spacetime
the FKKS Ansatz describes all possible modes.
The quasinormal modes for the FKKS
Ansatz
in Kerr were then
recalculated and found to yield excellent agreement with previous
approaches~\cite{Percival:2020skc}. The investigation of the degrees
of freedom covered by the FKKS Ansatz
in spinning geometries can be
performed either in a small spin approximation or numerically, whereas
analytically, this investigation is arduous due to the complexity of
the system comprised of partial differential equations and a
polynomial equation.

In this paper, we analyze Proca fields in the Schwarzschild-AdS
spacetime through both the VSH and the FKKS techniques, i.e., we
investigate the degrees of freedom and quasinormal modes through both
methods.
In the computation of the quasinormal modes, we use reflective
boundary conditions at infinity and pure incoming wave boundary
conditions at the event horizon.
The analysis in the VSH method has the advantage that both
the electric and the magnetic sectors appear in a natural way.
The analysis in the FKKS method has the advantage that the
polarizations in the electric mode sector decouple promptly.
To do this we use 
 a useful
numerical procedure
set up by
Horowitz and Hubeny \cite{Horowitz:2000}
to study perturbations for scalar fields 
in
Schwarzschild-AdS spacetimes and 
to find their quasinormal mode frequencies,
see also
\cite{cl}. An extension of this method 
which substitutes the
number coefficients by matrix coefficients
will be performed by us here,
see also \cite{Delsate:2011qp}. 
For a review of all these treatments and
an overview of the FKKS Ansatz see the 
thesis~\cite{thesis}.

The paper is organized as follows. In
Sec.~\ref{sec:EinsteinProcafieldequations}, we present the general
Einstein-Proca field equations and specialize them for a fixed
background geometry
with a negative cosmological constant.
In Sec.~\ref{sec:ProcafieldSchwADS}, we
separate the Proca equations for a Schwarzschild-AdS spacetime using
the VSH method and compute its quasinormal modes. In
Sec.~\ref{sec:FKKS}, we present the FKKS method adapted to the
Kerr-AdS spacetime, take the nonrotating limit to the
Schwarzschild-AdS spacetime and compute its quasinormal modes. In
Sec.~\ref{sec:compare}, we compare the quasinormal modes given by
the VSH and FKKS methods. In Sec.~\ref{conc}, we
conclude. Geometric units $G=c=1$ are used throughout the paper.

\section{Einstein-Proca field equations
with a negative cosmological constant: Fixed background}
\label{sec:EinsteinProcafieldequations}
The action for the minimally coupled Einstein-Proca system is 
given by
\begin{equation}
S = \int \dd[4]x \sqrt{-g}(\mathcal{L}_{EH} - \mathcal{L}_A)\,,
\end{equation}
where $g$ is the determinant of the metric $g_{ab}$,
\begin{equation}
\mathcal{L}_{EH}=\frac{R - 2\Lambda}{16\pi}
\end{equation}
is the
Einstein-Hilbert Lagrangian density
with $R=R_{ab}g^{ab}$, and
$R_{ab}$  the Ricci tensor, both built out of
the Riemann tensor ${R^a}_{bcd}$ made of the metric
and its first and second derivatives.
$\Lambda$ is the cosmological constant, with which we can
define a characteristic length $l = \sqrt{\frac{3}{\abs{\Lambda}}}$,
and 
\begin{equation}
\mathcal{L}_A = \frac{F_{ab}F^{ab}}{4} + \frac{\mu^2}{2} A_a A^a\,,
\end{equation}
is the Proca Lagrangian density with $F_{ab}$, given by $F_{ab} =
\nabla_{a} A_{b} - \nabla_{b} A_{a}$ being the Proca field strength,
$A_a$ being the Proca vector potential, and $\mu$ the field mass.
It follows from
the Euler-Lagrange equations
for the metric field
$g_{ab}$
that the
Einstein equation
must be obeyed, i.e., 
\begin{align}
  G_{ab} + \Lambda g_{ab} = 8\pi T_{ab}\,,
  \label{e1}
\end{align}
where $G_{ab}$ is the Einstein tensor
defined
as $G_{ab}=R_{ab}- \frac{1}{2} g_{ab} R$, 
and $T_{ab}$ is the energy-momentum tensor
given by
\begin{align}
T_{ab} = F_{ac}F_{bd}g^{cd}-
\frac14 g_{ab}F_{cd}F^{cd}
+ \mu^2 \left(A_a A_b-\frac12 g_{ab}
A_c A^c\right)
\,.
\label{eq:stressenergyVector}
\end{align}
It follows from
the Euler-Lagrange equations
for the Proca field 
$A_a$ that the
Proca equation
must be obeyed, i.e.,
\begin{align}
\nabla_b F^{ab} + \mu^2 A^a = 0 \,.
  \label{eq:Procaeqvacuum0}
\end{align}
The tensor field $F_{ab}$
also obeys
the internal equations $\nabla_{[a}F_{bc]} = 0$.

The Einstein-Proca system
consists of nonlinear coupled partial
differential equations, Eqs.~(\ref{e1})-(\ref{eq:Procaeqvacuum0}).
We now want to study the perturbations of the
$A_a$ field in a fixed background
with  negative cosmological
constant, $\Lambda<0$. In other words, we consider
a perturbative expansion of the field equations when
the vector field and its derivatives are small.
In this case, from Eq.~(\ref{e1}) one has the vacuum Einstein equation,
namely,
\begin{align}
G_{ab} - \frac{3}{l^2} g_{ab} = 0\,.
\label{eins2}
\end{align}
Now, Eq.~(\ref{eins2})
obeys the field Bianchi identities,
$\nabla_a\left(G^{ab} - \frac{3}{l^2} g^{ab} \right)=0$.
Although to first order
the stress-energy tensor $T_{ab}$ is zero, in second order it is not;
it must obey the conservation equation
\begin{align}
\nabla_a T^{ab} = 0\,,
\label{eq:tabconserved}
\end{align}
where $T_{ab}$ is the
Proca energy-momentum tensor given in
Eq.~(\ref{eq:stressenergyVector}).
For the Proca equation given in 
Eq.~(\ref{eq:Procaeqvacuum0}),
one can make use of the commutation
relations for covariant derivatives
that involve the Riemann tensor
to yield
$g^{cd}\nabla_c\nabla_d A^a - \mu^2 A^a -
R^{a}_{\,\,d}A^d = 0 $,
where
use of
the Bianchi identity for the Proca field was
also made.
In a background spacetime with a
negative cosmological constant,
we have from
Eq.~(\ref{eins2}) that $R_{ab}=- \frac{3}{l^2} g_{ab}$.
So the Proca equation
in a fixed negative cosmological constant background
becomes 
\begin{align}
g^{cd}\nabla_c\nabla_d A^a
- \left(\mu^2 - \frac{3}{l^2}\right)A^a = 0 \,.
\label{eq:Procaeqvacuum}
\end{align}
The Bianchi identity for the Proca  $A^a$ field is 
clearly
\begin{align}
\nabla_aA^a = 0\,,
\label{eq:LorenzCondition}
\end{align}
i.e., its divergence is zero.  Note that for the massless case,
$\mu=0$, the Bianchi identity given in Eq.~(\ref{eq:LorenzCondition})
is the Lorenz gauge condition that
may or may not be imposed, indeed in
this case the field equations are invariant under the gauge
transformation~$ A_a \rightarrow A_a +
\nabla_a\chi$, where~$\chi$ is a
scalar obeying the Klein-Gordon equation, i.e. $\nabla_a \nabla^a \chi
= 0$.

Thus, to study the perturbations of the Proca field $A_a$ with a fixed
spacetime background, and thus a fixed metric, one uses
Eq.~\eqref{eq:Procaeqvacuum} with the help of
Eq.~\eqref{eq:LorenzCondition}.  Of interest is the case in which the
cosmological constant is negative and the background is a
Schwarzschild-AdS black hole.  We apply two methods to solve the Proca
set of equations given in
Eqs.~\eqref{eq:Procaeqvacuum} and \eqref{eq:LorenzCondition},
the VSH and
the FKKS methods.

\section{Proca field perturbations in Schwarzschild-AdS:
The VSH method}
\label{sec:ProcafieldSchwADS}

\subsection{Schwarzschild-AdS metric}

The spacetime considered
here is
the Schwarzschild-AdS spacetime, whose line element can be written in
spherical $(t,r,\theta,\phi)$ coordinates as
\begin{equation}
ds^2 = - f(r) dt^2 + \dfrac{dr^2}{f(r)}
+ r^2 (d\theta^2 + \sin^2\theta d\phi^2)\,,
\label{eq:fSchwADS}
\end{equation}
where the function $f(r)$ is
\begin{equation}
f(r) = 1 - \frac{2M}{r}+\frac{r^2}{l^2} \,,
\label{eq:fSchwADS2}
\end{equation}
and~$M$ is the spacetime mass.
The Schwarzschild-AdS metric describes a
vacuum
spherically symmetric spacetime with a negative cosmological
constant. It is a static black hole solution with an 
event horizon located at $r_+$ given by
the positive root of the equation $f(r)=0$, i.e.,
\begin{equation}
r_+^3+r_+l^2 - 2Ml^2=0\,.\label{eq:fSchwADShorizon}
\end{equation}

\subsection{Perturbations in the Proca field,
separation of variables, and the VSH method}

\subsubsection{General case}

In the Schwarzschild-AdS spacetime, the Proca
equations~\eqref{eq:Procaeqvacuum} can be separated using vector
spherical harmonics
which can be obtained by the sum of a
spin~$s=1$ with angular momentum~$\ell$
as it is well known
from any modern quantum mechanics textbook.
This  vector
spherical harmonics method, or VSH method, was first 
used in~\cite{Ruffini:1973} to separate the Maxwell equations 
in static geometries.
In the
VSH method the following Ansatz
is  assumed for the field
$A_a$,
\begin{align}
  A_{a} = \frac{1}{r}\sum_{i=0}^3\sum_{l m} c_i u_{(i)}(t,r)
  Z_{a}^{(i)\ell m}(\theta,\phi)\,,
  \label{eq:SchwParam}
\end{align}
where~$c_0 =1 $, $c_1 = 1$, $c_2
=\frac{1}{\left[\ell(\ell+1)\right]^{\frac{1}{2}}}$,
$c_3
=\frac{1}{\left[\ell(\ell+1)\right]^{\frac{1}{2}}}$, 
the $u_{(i)}$, with $i=0,1,2,3$,  are functions of $t$ and $r$
that should be written as
$u_{(i)}(t,r)\equiv u_{(i)}^{\ell m}(t,r)$,
but the indices $\ell$ and $m$ have been suppressed to not
overcrowd the notation,
and the $Z_{a}^{(i)\ell m}$ are given by
\begin{align}
    Z_{a}^{(0)\ell m} &= (1,0,0,0) Y^{\ell m}\,,\\
    Z_{a}^{(1)\ell m} &= (0,\dfrac1f,0,0)Y^{\ell m}\,,\\
    Z_{a}^{(2)\ell m} &= \frac{r}{\sqrt{\ell(\ell+1)}}
    (0,0,\p_{\theta},\p_{\phi})Y^{\ell m}\,,\\
    Z_{a}^{(3)\ell m} &= \frac{r}{\sqrt{\ell(\ell+1)}}
    \Bigg(0,0,\frac{\p_{\phi}}{\sin\theta},
    -\sin\theta\p_{\theta}\Bigg)Y^{\ell m}\,,
\end{align}
where $Y^{\ell m}$ are the spherical harmonics,
with $\ell$ being the principal number and
$m$  the azimuthal number.

Indeed, by inserting the Ansatz
given in Eq.~(\ref{eq:SchwParam})
into the Proca
equations~\eqref{eq:Procaeqvacuum},
one finds after rearrangement
the following system of equations for the
functions~$u_{(i)}$,
\begin{align}
  &{\hat{\mathcal{D}}}^2u_{(0)} +
  (\p_r f)(\dot{u}_{(1)} - u'_{(0)}) = 0 \,,
  \label{eq:SchwEM1}\\
&  {\hat{\mathcal{D}}}^2 u_{(1)} + \frac{2f}{r^2}
  \Big(1 - \frac{3M}{r}\Big) (u_{(2)} - u_{(1)})=0\,,
  \label{eq:SchwEM2better}\\
&{\hat{\mathcal{D}}}^2 u_{(2)}
  + \Bigg[\frac{2f \ell(\ell+1)}{r^2}u_{(1)}\Bigg] = 0\,,
  \label{eq:SchwEM3}\\
  &{\hat{\mathcal{D}}}^2 u_{(3)} = 0\,,\label{eq:SchwEM4}
\end{align}
where~$\dot{u}_{(i)} = \dfrac {\partial{u_{(i)}}} {\partial t}$,
$u'_{(i)} = \dfrac {\partial{u_{(i)}}} {\partial r_*}$
with  $r_*$ being defined by
$\dfrac{dr_*}{dr} = \dfrac1f$, and where
${\hat{\mathcal{D}}}^2$ is shorthand for
\begin{equation}
  {\hat{\mathcal{D}}}^2  = - \p_t^2 + \p_{r_*}^2
  - f\Big[\frac{\ell(\ell+1)}{r^2} + \mu^2\Big]\,.
\label{d2}
\end{equation}
It must be noted that $u_{(0)}$, $u_{(1)}$, and $u_{(2)}$  describe
the electric modes  since, under parity
transformations,
$Z^{(0)}_a$, $Z^{(1)}_a$, and  $Z^{(2)}_a$ gain a
factor of~$(-1)^{\ell}$,
whereas $u_{(3)}$
describes the magnetic modes
  since, under parity
transformations,
$Z^{(3)}_a$ gains
a factor of~$(-1)^{\ell+1}$.
All this follows the notation given in~\cite{Rosa:2012},
where the Proca equations for a pure
Schwarzschild background have been
presented.

Furthermore, inserting
the Ansatz in Eq.~(\ref{eq:SchwParam})
into the Bianchi identity
given in Eq.~(\ref{eq:LorenzCondition}),
$\nabla^a A_a =0$,
one obtains
\begin{equation}
  \frac{1}{r f}\Bigg[u'_{(1)} - \dot{u}_{(0)}
  + \frac{f}{r} (u_{(1)} - u_{(2)}) \Bigg] = 0\,.
  \label{eq:LorenzConditionSchw}
\end{equation}
Equation~~\eqref{eq:LorenzConditionSchw}
was used to find Eq.~(\ref{eq:SchwEM2better}).
Indeed, the Ansatz given in Eq.~(\ref{eq:SchwParam})
when put into Eq.~\eqref{eq:Procaeqvacuum}
leads directly to 
$\hat{\mathcal{D}}u_{(1)} + (\p_r f) (\dot{u}_{(0)} - u'_{(1)})
+ \frac{2 f^2}{r^2}(u_{(2)} - u_{(1)})  = 0$,
which upon using
Eq.~(\ref{eq:LorenzConditionSchw})
yields Eq.~(\ref{eq:SchwEM2better}).

Therefore, the system consisting
of Eqs.~\eqref{eq:SchwEM1}-\eqref{eq:SchwEM4},
taking into account the definition given in Eq.~\eqref{d2},
determines the solution. The Bianchi identity,
Eq.~\eqref{eq:LorenzConditionSchw}, also helps in the
determination of the solution. For example,
the static part
of~$u_{(0)}$
 must be obtained from
Eq.~\eqref{eq:SchwEM1}, but the 
dynamical part of~$u_{(0)}$ can be described by the  Bianchi identity
Eq.~\eqref{eq:LorenzConditionSchw}.
Equations~\eqref{eq:SchwEM2better} and~\eqref{eq:SchwEM3},
for~$u_{(1)}$ and~$u_{(2)}$,
are coupled together, whereas
Eq.~\eqref{eq:SchwEM4}, for~$u_{(3)}$, is
decoupled.

We shall assume that the time dependence of the
functions $u_{(0)}$,
$u_{(1)}$,
$u_{(2)}$, and 
$u_{(3)}$ goes as
$e^{-i\omega t}$. In this case, the system
given by Eqs.~\eqref{eq:SchwEM1}-\eqref{eq:SchwEM4}
can be treated as an eigenvalue
problem to $\omega$. We classify the eigenvectors of the system
according to the three degrees of freedom, i.e.,
the three polarizations, of the
Proca vector $A_a$.
These three polarizations consist of one scalar-type
polarization and two vector-type polarizations.
The electric modes of $A_a$, characterized by $u_{(0)}$,
$u_{(1)}$, and
$u_{(2)}$, possess one
scalar-type polarization and one
vector-type polarization.
The scalar-type
polarization of $A_a$ has
a behavior similar to a scalar
field, and can be picked up to the higher $\ell$ modes
from the $\ell = 0$ mode,
which has solely scalar-type polarization.
Moreover, setting the Proca mass $\mu$
to zero,~$\mu = 0$, 
the scalar-type polarization becomes nonphysical,
more precisely, at
the massless limit it can be removed by the gauge freedom.
The vector-type
polarization of $A_a$ of the electric modes
are then picked up by exclusion, i.e., they are the ones that
are not scalar-type.
Since the system in the
electric mode sector is coupled, it is not trivial to obtain the
scalar-type and the vector-type polarization of $A_a$ in terms of the
functions $u_{(1)}$ and $u_{(2)}$.
The magnetic modes of $A_a$, characterized by $u_{(3)}$,
possess the remaining 
vector-type polarization.

As an addendum, we note that
it is possible
to decouple the pair of equations for $u_{(1)}$ and
$u_{(2)}$ which are also present in the
pure Schwarzschild case,
see~\cite{Rosa:2012}.
Equation~\eqref{eq:SchwEM3} can be used to
find~$u_{(1)}$ as a function of~$u_{(2)}$, i.e.,
$ u_{(1)} = -
\frac{r^2 \hat{\mathcal{D}}^2u_{(2)}}{2 f \ell (\ell + 1)}$.
Substituting this expression in Eq.~\eqref{eq:SchwEM2better}, a
decoupled equation for~$u_{(2)}$ can be found,
$\mathcal{H}\left(r u_{(2)}\right) = 0$, where
$\mathcal{H} = \hat{\mathcal{D}}^2
\Big[\frac{1}{f}\hat{\mathcal{D}}_1^{\,2} \Big] -
2 f (\partial_r f) \mu^2$ and
${\hat{\mathcal{D}}_1}^{\,2} = - \partial_t^2 +
\partial_{r_*}^2
  - f\Big[\frac{\ell (\ell + 1)}{r^2} + \mu^2
    + \frac{\partial_r f}{r}\Big]$.
Thus, one can decouple the pair of equations
in this way by paying the
price of increasing the order of the partial
differential equations. This happens
since $u_{(2)}$ must contain the scalar-type and the
vector-type polarizations. In the massless
limit, $\mu=0$,
the operator $\mathcal{H}$ factorizes, becoming the product
between the operators $\hat{\mathcal{D}}^2$ and
$f^{-1}\hat{\mathcal{D}}_1^{\,2}$. One must notice that in this limit,
the scalar-type polarization can be removed by the gauge freedom,
meaning $u_{(2)}$ will contain both
a spurious degree of freedom and the physical
vector-type polarization associated with the electric modes.
It can be shown that the scalar~$\Psi =
f^{-1}\mathcal{D}_1^{\,2}(r u_{(2)})$ is related to the field
strength
tensor $F_{ab}$, thus
in the massless case it describes appropriately the vector-type
polarization related to the electric modes without the spurious degree
of freedom. Moreover, the equation $\mathcal{H}(ru_{(2)}) = 0$
indicates that $\Psi$ will satisfy the same equation as~$u_{(3)}$,
which means the two vector-type polarizations of the massless vector
field become degenerate. For the massive case, the factorization of
$\mathcal{H}$ does not appear to be possible, which makes the
analytical decoupling difficult.

\subsubsection{Monopole case}
The monopole case~$\ell=0$ for the massive vector field simplifies the
system considerably.
Only the functions 
$u_{(0)}$ and $u_{(1)}$ survive, 
the functions~$u_{(2)}$ and~$u_{(3)}$ vanish
since~$Y^{00}$ is a constant.
The equation for the function~$u_{(0)}$
can be obtained from
Eq.~\eqref{eq:SchwEM1} together with Eq.~\eqref{d2}, 
to give 
\begin{align}
  -\ddot{u}_{(0)}+u_{(0)}'' + (\partial_r f)(\dot{u}_{(1)}  - u_{(0)}')
  - f \mu^2 u_{(0)} =  0\,.
  \label{eq:SchwMonopole1st}
\end{align}
The equation for~$u_{(1)}$ can be
obtained from
Eq.~\eqref{eq:SchwEM2better} together with Eq.~\eqref{d2}, i.e., 
\begin{align}
  u''_{(1)} - \ddot{u}_{(1)} - f\Big[\mu^2 + \frac{2}{r^2}
    \Big(1- \frac{3 M}{r} \Big) \Big]u_{(1)} = 0\,.
  \label{eq:SchwMonopole1dyn}
\end{align}
The Bianchi identity given in Eq.~\eqref{eq:LorenzConditionSchw}
is now
\begin{align}
{\dot{u}}_{(0)}
= u'_{(1)} + \frac{f}{r} u_{(1)}\,.
\label{bi2}
\end{align}

Note that for the~$\ell=0$
case the function~$u_{(0)}$
could be written as 
$
u_{(0)}=u_{(0)\text{s}}(r)+
u_{(0)\text{t}}(t,r)
$,
where
$u_{(0)\text{s}}(r)$
is the static part
of $u_{(0)}$
and  can be obtained
directly from
Eq.~\eqref{eq:SchwMonopole1st}, and
$u_{(0)\text{t}}(t,r)$ is 
the dynamic part of~$u_{(0)}$ and
can be obtained directly from the
Bianchi identity given in Eq.~(\ref{bi2}).
The function $u_{(1)}$ is taken
from Eq.~\eqref{eq:SchwMonopole1dyn}.
We are not considering the static part which would give a static
spherically symmetric second-order perturbed
Schwarzschild-AdS geometry with a new Proca gravitational term with the
corresponding first-order Proca field term, rather than the
Schwarzschild-AdS background geometry we are working with.  By taking
the zero-mass field limit, the static part would give
a static
spherically symmetric second-order perturbed
Schwarzschild-AdS geometry with a new Maxwell gravitational term, i.e.,
a second-order
Reissner-Nordstr\"om-AdS  geometry, with the corresponding
first-order Maxwell field.

\subsection{Quasinormal modes of Proca
in Schwarzschild-AdS in the VSH method}
\label{sec:quasinormal}


Having found the equations obeyed by the Proca field in
a Schwarzschild-AdS background, namely
Eqs.~(\ref{eq:SchwEM1})-(\ref{eq:SchwEM4}) for $u_{(i)}$
with $i=0,1,2,3$, 
we can now study the quasinormal modes
of a Schwarzschild-AdS black hole for a Proca field.
The quasinormal modes are defined as
solutions that solve the equations of motion given in
Eqs.~(\ref{eq:SchwEM1})-(\ref{eq:SchwEM4}) with
boundary conditions such that at the event horizon there are only 
purely incoming waves and at infinity the Proca field is zero, i.e.,
$u_{(i)} \rightarrow 0$ at infinity, with
$i=0,1,2,3$.
The analysis of the system is concluded by integrating the
equations.
To find the quasinormal modes
and the quasinormal frequencies in this spacetime, we
implement the 
numerical procedure used in
\cite{Horowitz:2000},
see also \cite{Cardoso:2001}. 
One can write $u_{(i)}$, with $i=0,1,2,3$, as 
\begin{align}
 u_{(i)}= 
U_{(i)} e^{-i (t + r_*)\omega} \,,\label{eq:Horowitzsol0}
\end{align}
where $\omega$ is a frequency
and the $U_{(i)}$ are functions of $r$.
This transformation is useful since it expresses explicitly the
behavior of $u_{(i)}$ as an incoming wave at the event horizon.
Assuming
analyticity, it is possible to write
every $U_{(i)}$
as 
an expansion series around the horizon $r_+$, 
\begin{align}
U_{(i)} =   \sum_{n=0}^{\infty} a_{(i)n}
(x - x_+)^n\,,\label{eq:Horowitzsol}
\end{align}
where
the $a_{(i)n}$ are expansion coefficients,
$x = \frac{1}{r}$, and $x_+ = \frac{1}{r_+}$.

\begin{table*}
\centering
        \begin{subtable}{1\textwidth}
    \centering
    \begin{tabular}{|c|c|c|}
        \hline
                        & $r_+=l$ & $r_+ = 100 l$\\\hline
        $\mu l$ &   $\omega l$ (VSH) & $\omega l$ (VSH)
	                            \\ \hline
        $0.01$ & $3.331 -  2.489\,\, i$ &   $184.968 - 266.394\,\, i$  \\ \hline
        $0.10$ & $3.339 - 2.500\,\, i$ &   $185.604 - 267.461\,\, i$ \\ \hline
        $0.20$ & $3.362 -  2.531\,\, i$ &    $187.452 - 270.612\,\, i$ \\ \hline
        $0.40$ & $3.446 -  2.645\,\, i$ &  $193.925 - 282.119\,\, i$ \\ \hline
        $0.50$ & $3.501 - 2.722\,\, i$ & $198.077 - 289.799\,\, i$ \\ \hline
        \end{tabular}
        \caption{Scalar-type polarization.\label{tab:QNMmBmVSHSAdS}}
\end{subtable}
\vskip 0.75cm
\begin{subtable}{1\textwidth}
\centering
    \begin{tabular}{|c|c|c|}
        \hline
                        & $r_+=l$& $r_+ = 100 l$\\\hline
        $\mu l$ &   $\omega l$ (VSH) & $\omega l$ (VSH) \\ \hline
        $0.01$ & $1.554 - 0.542 \,\,i$ &   $0 - 149.984\,\, i$\\ \hline
        $0.10$ & $1.557 - 0.552\,\, i$ &   $0 - 152.099\,\, i$  \\ \hline
        $0.20$ & $1.568 -  0.583\,\, i$ &    $0 - 158.432\,\, i$\\ \hline
        $0.30$ & $1.585 -  0.633\,\, i$ &  $0 - 168.817\,\, i$ \\ \hline
        $0.40$ & $1.607 -  0.699\,\, i$ &  $0 - 183.291\,\, i$ \\ \hline
        $0.50$ & $1.634 - 0.777\,\, i$ & $0 - 202.684\,\, i$ \\ \hline
        \end{tabular}
        \caption{Vector-type polarization.
        \label{tab:QNMmBpVSHSAdS}}
        \end{subtable}\vspace{1mm}
\caption{
Quasinormal mode frequencies $\omega l$ of the Proca field electric
modes $u_{(0)}$, $u_{(1)}$, and $u_{(2)}$ with~$\ell = 1$,
using the VSH method in Schwarzschild-AdS for~$r_+ =
l$ and $r_+ = 100l$ and for several values of the Proca field
mass~$\mu l$. (a) The frequencies of the scalar-type polarization 
of the electric modes are displayed. 
(b) The frequencies of the vector-type polarization 
of the electric modes are displayed.
\label{tab:QNMVSH}}
  \label{table1u1u2u3}
\end{table*}

\begin{table*}
 \centering
    \centering
    \begin{tabular}{|c|c|c|}
        \hline
                                & $r_+=l$ & $r_+ = 100 l$\\\hline
        $\mu l$ &   $\omega l$ (VSH) & $\omega l$ (VSH)
	                            \\ \hline
        $0.01$ & $2.163 -  1.699\,\, i$ &   $0 - 150.069\,\, i$ \\ \hline
        $0.10$ & $2.171 - 1.710\,\, i$ &   $0 - 152.187\,\, i$ \\ \hline
        $0.20$ & $2.193 -  1.743\,\, i$ &   $0 - 158.526\,\, i$ \\ \hline
        $0.30$ & $2.228 -  1.795\,\, i$ &   $0 - 168.922\,\, i$ \\ \hline
        $0.40$ & $2.273 -  1.863\,\, i$ &   $0 - 183.419\,\, i$ \\ \hline
        $0.50$ & $2.327 - 1.944\,\, i$ &   $0 - 202.860\,\, i$ \\ \hline
    \end{tabular}
    \caption{
Quasinormal mode frequencies $\omega l$ of the Proca field magnetic
modes $u_{(3)}$ with~$\ell = 1$, using 
the VSH method in Schwarzschild-AdS for~$r_+ =
l$ and $r_+ = 100l$ and for several values of the Proca field
mass~$\mu l$. Magnetic modes only have vector-type polarization.
      \label{tab:u4}}
\end{table*}

The functions $u_{(0)}$, $u_{(1)}$, and
$u_{(2)}$ give the electric modes.
For $u_{(0)}$, the equation to be solved
is given in 
Eq.~(\ref{eq:SchwEM1}), and one sees it is coupled to 
the equation for
$u_{(1)}$,
Eq.~(\ref{eq:SchwEM2better}).
The variables 
$u_{(1)}$ and $u_{(2)}$ are also coupled, see
Eqs.~(\ref{eq:SchwEM2better})
and (\ref{eq:SchwEM3}),
and this means
that  careful treatment
is required to solve them.
The strategy that we follow here is
to determine $u_{(1)}$ and $u_{(2)}$ from 
Eqs.~(\ref{eq:SchwEM2better}) and~(\ref{eq:SchwEM3})
and then use the Bianchi identity to determine 
Eq.~(\ref{eq:LorenzConditionSchw}) to determine
$u_{(0)}$.
Since the Horowitz-Hubeny
numerical procedure \cite {Horowitz:2000}
was designed for decoupled
equations, an extension is needed
for this case. Hence we substitute the
number coefficients by matrix coefficients,
see also~\cite{Delsate:2011qp,thesis}.
Let us first define 
the following  polynomials by
\begin{align}
  &s(x) = \frac{x^4f(x)}{x-x_+}\,\label{eq:poly}\,,\\
  &t(x) = x^2 \p_x\Big(
  x^2f(x)\Big) + 2i \omega x^2\,,\label{eq:poly1}\\
  &u(x) = - (x-x_+)\Big[x^2 \ell(\ell + 1) +
  \mu^2 \Big] \,,\label{eq:poly2}
\end{align}
where $f(x)$ is $f(r)$ of Eq.(\ref{eq:fSchwADS2})
transformed to the variable $x$,
i.e.,
\begin{align}
f(x)=
\frac1{x^2}\left(
\frac1{l^2} + x^2 - 2Mx^3\right)\,,
\end{align}
and the matrix $\bm{K}$ by
\begin{align}
    \bm{K} = (x-x_+)
    \begin{bmatrix}
    - 2 x^2 (1- 3 M x) & 2 x^2 (1- 3 M x) \\
    2 x^2 \ell(\ell + 1) & 0 \\
    \end{bmatrix}\,. 
    \label{matrixK}
\end{align}
We then substitute Eq.~\eqref{eq:Horowitzsol0} into
Eqs.~\eqref{eq:SchwEM2better} and~\eqref{eq:SchwEM3}, finding a matrix
equation for the $U_{(i)}$ given by
\begin{align}
  &(x - x_+) s(x)\p_x^2 \bm{U} + t(x)\p_x \bm{U}
   +\frac{u(x)}{x-x_+} \bm{U} \nonumber\\
  &+\frac{1}{x-x_+}\bm{K}\cdot\bm{U} = 0\,,
\label{theequation1}
\end{align}
where the matrix $\bm{U}$ is defined in a natural
way by
\begin{align}
   \bm{U} = \begin{bmatrix} 
    U_{(1)} \\
    U_{(2)}
    \end{bmatrix}\,.
    \label{Umatrixoriginal}
\end{align}
Notice that in the polynomials above, there is only
a linear dependence in $\omega$. This happens because the second time
derivative and the second $r_*$ derivative have different signs in
Eq.~\eqref{d2}, and when applying each of the derivatives to the
exponential in Eq.~\eqref{eq:Horowitzsol0} the term in $\omega^2$
cancels.  
Now, to solve the problem we have to
expand the matrix $\bm{U}$ as
it was done in Eq.~\eqref{eq:Horowitzsol},
i.e.,
\begin{align}
\bm{U}(x) = \sum_{n=0}^{\infty} {\bm{a}_n}(x-x_+)^n\,,
\label{Umatrixexpansion}
\end{align}
where
$\bm{a}_n$ for the $a_{(i)n}$, $i=1,2$, is given by
\begin{align}
    \bm{a}_n = \begin{bmatrix}
    a_{(1)n} \\
    a_{(2)n}\,
    \end{bmatrix}.
\label{matrix1}    
\end{align}
It is helpful to write
\begin{align}
    \bm{a}_n = \bm{M}_n \bm{a}_0 \,\,
\label{matrix2}    
\end{align}
for some $2\times 2$ matrix $\bm{M}_n$
that has to be found,
with the obvious definition that $\bm{M}_0$
is the identity matrix $\bm{I}$, i.e.,
$\bm{M}_0 = \bm{I} =
\begin{bmatrix}
1 & 0 \\
0 & 1 
\end{bmatrix}$.
Note now 
that the polynomials
defined in Eqs.~(\ref{eq:poly})-(\ref{eq:poly2})
can be expanded around~$x_+$ so that it is possible
to write
\begin{equation}
s(x) = \sum_{j=0}^{\infty} s_j
(x-x_+)^j\,,
\label{s}
\end{equation}
\begin{equation}
t(x) = \sum_{j=0}^{\infty} t_j
(x-x_+)^j\,,
\label{t}
\end{equation}
\begin{equation}
u(x) = \sum_{j=0}^{\infty} u_j
(x-x_+)^j\,,
\label{u}
\end{equation}
where $s_j$, $t_j$, and $u_j$ are expansion coefficients that vanish
for $j$ higher than some value, since $s(x)$, $t(x)$, and $u(x)$
are finite polynomials. We also can expand $\bm{K}$ in
Eq.~(\ref{matrixK})
as
\begin{equation}
\bm{K}(x) = \sum_{j=0}^{\infty} {\bm{K}}_j(x-x_+)^j\,,
\end{equation}
where ${\bm{K}}_j$ are the expansion coefficients which also vanish
for $j$ higher than some value, since the components of the matrix
$\bm{K}$ are finite polynomials.  Equation~(\ref{theequation1}), which
is equivalent to the
Proca equations for~$u_{(1)}$ and~$u_{(2)}$,
is then reduced to the
recurrence relation
\begin{align}
  \bm{M}_n = - \frac{1}{P_n}
  \sum_{j=0}^{n-1}\bm{V}_{nj}\cdot\bm{M}_j\,,
  \label{Mn}
\end{align}
where
\begin{align}
 \bm{V}_{nj} = \big[j(j-1 )s_{n-j} + j t_{n-j} + u_{n-j}\big]
  \bm{I} + \bm{K}_{n-j}\,\label{eq:recurrencecompl}
\end{align}
and
\begin{equation}
  P_n =  n(n-1) s_0 + n t_0 \,. \label{eq:HorowitzHubeny} 
\end{equation}
The quasinormal mode frequencies for
the functions~$U_{(1)}$ and~$U_{(2)}$
that appear in Eq.~\eqref{eq:Horowitzsol} and are 
put in matrix form in  Eq.~\eqref{Umatrixoriginal}
can be obtained by imposing that the series
appearing in Eq.~\eqref{Umatrixexpansion}
vanishes  at~$x\rightarrow0$,
i.e., $r\rightarrow +\infty$.
More specifically, from Eqs.~\eqref{Umatrixoriginal}-\eqref{matrix2},
the series given in Eq.~\eqref{Umatrixexpansion} vanishes
at~$x\rightarrow 0$ if either $\bm{a}_0 = 0$, which means the series
vanishes everywhere trivially, or $\sum_{j=0}^\infty \bm{M}_j
(-x_+)^j$ is singular, which means the determinant of the matrix
resulting from the sum is zero.  Thus, discarding the trivial
solution, the boundary condition is satisfied when
\begin{align}
    \det\Bigg(\sum_{j=0}^N \bm{M}_j (-x_+)^j\Bigg) = 0\,,
\end{align}
where $N$ is in principle infinite.
Then, for $u_{(0)}$
the quasinormal mode frequencies
are directly determined through the
Bianchi identity,
Eq.~(\ref{eq:LorenzConditionSchw}).
Of course, the quasinormal frequencies
for $u_{(0)}$,
$u_{(1)}$, and
$u_{(2)}$ are the same,
they are the electric modes.  We do not present the modes for $\ell =
0$, but they were calculated and agree with~\cite{Konoplya:2006}.  The
modes for $\ell = 1$ are shown in the Table~\ref{table1u1u2u3}, in
particular the scalar-type polarization is shown in
Table~\ref{tab:QNMmBmVSHSAdS} and the vector-type polarization is
shown in Table~\ref{tab:QNMmBpVSHSAdS}, where $N=40$ was taken, and we
set $r_+$ and $\omega$ in units of $l$ and $l^{-1}$, respectively.  
In the tables we wrote the number 0 which in the
numerical procedure we use means a real number that is very close to
zero. This hints that the quasinormal modes are purely imaginary.

The function $u_{(3)}$
gives the magnetic modes.
For $u_{(3)}$, 
Eq.~\eqref{eq:SchwEM4} can be written with
the help of
Eq.~\eqref{eq:Horowitzsol0} and
with $x = \frac{1}{r}$ and $x_+ = \frac{1}{r_+}$
as
\begin{align}
  \Bigg[(x - x_+) s(x)\p_x^2  + t(x)\p_x
    + \frac{u(x)}{x-x_+} \Bigg]U_{(3)} = 0\,,
    \label{eq:Horowitzeq}
\end{align}
where the polynomials
$s(x)$,
$t(x)$, and
$u(x)$
are in Eqs.~\eqref{eq:poly}-\eqref{eq:poly2}.
Now to solve the problem we have to
expand  $U_{(3)}$ as in Eq.~\eqref{eq:Horowitzsol},
i.e.,
\begin{align}
U_{(3)} =   \sum_{n=0}^{\infty} a_{(3)n}
(x - x_+)^n\,,\label{eq:Horowitzsolu4}
\end{align}
where 
the $a_{(3)n}$ are expansion coefficients.
Using the expansions for $s(x)$,
$t(x)$, and
$u(x)$
given in
Eqs.~\eqref{s}-\eqref{u},
one finds that
Eq.~\eqref{eq:Horowitzeq} can then be reduced to the
recurrence relation,
\begin{align}
  &a_{(3)n} = - \frac{1}{P_n} \sum_{j=0}^{n-1}
  \Big(j(j-1) s_{n-j} + j t_{n-j} + u_{n-j} \Big)a_{(3)j}\,.
  \label{eq:recurrencesimple}
\end{align}
The quasinormal mode
frequencies
$\omega$ can be found by imposing that the series
vanishes at~$x\rightarrow 0$, i.e., $r\rightarrow +\infty$,
\begin{align}
\sum_{j=0}^N a_{(3)j} (-x_+)^j = 0\,,
\end{align}
with $N$ being again, in principle, infinite.
The quasinormal frequencies
for $u_{(3)}$, i.e.,
the magnetic mode frequencies,
are calculated numerically 
for $\ell=1$ where $N=40$
was taken, and are
displayed in the Table~\ref{tab:u4}. We set $r_+$ and
$\omega$ in units of $l$ and $l^{-1}$, respectively.

A comment 
on the  distinction of the polarizations 
of the electric modes $u_{(0)}$, 
$u_{(1)}$, and $u_{(2)}$ is in order.
Since Eqs.~\eqref{eq:SchwEM2better} and~\eqref{eq:SchwEM3}
cannot be decoupled trivially, the distinction of the modes for each
polarization is made by inference. From Table~\ref{table1u1u2u3}, the
electric mode polarizations, namely the scalar-type and vector-type,
 are difficult to distinguish at small $\frac{r_+}l$, 
e.g., $r_+=l$. A possible
method to distinguish them is to compare the frequencies at $\ell=0$,
where only the modes of the scalar-type polarization are present,
with the frequencies at $\ell=1$, as it is expected they have a higher
modulus for higher $\ell$. For large 
$r_+$, e.g., $r_+ = 100l$, the distinction is easier
since we can compare the frequencies in 
Table~\ref{table1u1u2u3} with the frequencies in
Table~\ref{tab:u4}, because
the modes in the vector-type polarizations have the
same behavior, namely,
negligible real frequency. The reason for this is that the mass of
the field is very small compared with the mass of 
the black hole, thus the effect
of the field mass on the vector-type modes is almost negligible.
They also
approach the values computed for the massless case, done
in~\cite{Cardoso:2001}, which were confirmed by our numerics.

\section{Proca field perturbations in Schwarzschild-AdS:
The FKKS  method}\label{sec:FKKS}

\subsection{Kerr-AdS metric and the principal tensor in Kerr-AdS}

We now employ another, very interesting, method
to find the quasinormal modes of the 
Proca field in a Schwarzschild-AdS
background. The method relies on symmetries for a rotating
body in general relativity, specifically on
the symmetries of the Kerr-NUT-AdS  and of the Kerr-NUT-dS.
To use this formalism in the Schwarzschild-AdS
background we display it for the Kerr-AdS putting zero NUT charge
from the start and then take the limit $a=0$ of the 
 Kerr-AdS.

Symmetries are important in the analysis of physical systems as they
can allow for separability of field equations or
the integrability of equations of motion for test
particles. 
There are explicit
symmetries and hidden symmetries.
The quantity of interest related to hidden symmetries is the principal
tensor~$h_{ab}$, a nondegenerate closed conformal Killing-Yano 2-form,
i.e.,
an antisymmetric tensor. The principal
tensor~$h_{ab}$ is  nondegenerate
when its
matrix representation  in any
coordinate system has maximal rank.
Denoting  $i$ as the imaginary unit, 
and denominating for convenience
the eigenvalues
of
$h^a_{\,\,b}$ in four-dimensional spacetime 
as~$\pm i x_A$, where~$A = 1,2$,
nondegeneracy implies that the~$x_A$ are functionally independent
and nonvanishing, i.e., the Jacobian matrix of $X = (x_1,x_2)$ is
nonsingular.

The principal tensor is able to generate the Killing
tower~\cite{Frolov:2008jr}, a set of
symmetries that allows the integration of the Hamilton-Jacobi and the
Klein-Gordon equations in spinning geometries.  It was shown
in~\cite{Houri:2008ng} that 
Kerr-NUT
with a cosmological constant, a family of solutions
of the Einstein vacuum equations valid in
four and higher dimensions, is the unique family of 
spacetimes with a
principal tensor, under the condition 
that the gradient of the eigenvalues of the 
principal tensor
are spacelike vectors, or
timelike via Wick rotation. 
Principal tensors are very interesting 
quantities. As reviewed in~\cite{Frolov:2017}, 
they have applications in
the Kerr-NUT-AdS
and 
Kerr-NUT-dS families,
as well as in another
set of spacetimes with Lorentzian signature 
with a principal
tensor built from
eigenvalues with null
gradient, an issue that it is far from being fully
understood, and  is currently an open problem
\cite{Frolov:2017whj}.

Here we are interested in a
particular case of the four-dimensional
Kerr-NUT-AdS spacetime which is
the four-dimensional Kerr-AdS spacetime.
In Boyer-Lindquist coordinates
$(t,r,\theta,\phi)$, the Kerr-AdS spacetime
has line element given by
\begin{align}
  ds^2 =
  &
  - \frac{\Delta_\Lambda}{\Sigma}
  \Big[dt - a\sin^2\theta d\phi \Big]^2\notag\nonumber\\
  &
  + \frac{\Delta_\theta \sin^2\theta }{\Sigma}
  \Big[a dt - (a^2 + r^2)d\phi \Big]^2\notag\nonumber\\
  &
  +\frac{\Sigma}{\Delta_\Lambda}dr^2
  + \frac{\Sigma}{\Delta_\theta}d\theta^2\,,
\label{eq:KerrAdSmetric}
\end{align}
where
$\Delta_\Lambda = \,\, r^2 - 2Mr + a^2
  + \frac{r^2}{l^2}(r^2 + a^2)$,
$\Sigma = \,\, r^2 + a^2 \cos^2\theta$,
$\Delta_\theta = 1 - \frac{a^2}{l^2}
  \cos^2 \theta$,
and $a$ is related to the angular momentum of the
black hole, $J$, by $J
= a M$, with 
$M$ being the spacetime mass. The principal tensor
$h_{b c}$ obeys the following equation in this
spacetime
$\nabla_a h_{b c} = 2 g_{a[b}\xi_{c]}$,
where $\xi_{b}$
is defined by $\xi_{b} = \frac{1}{3}\nabla^c h_{c b}$,
and square brackets mean antisymmetrization on the indices.
In addition, 
$h_{b c}$ obeys
the following
integrability conditions,
$\nabla^a \nabla^b h_{c d} =
  -\Big(R^a_{\,\,e}\delta^b_{[c}h^e_{\,\,d]}
  + \frac{1}{2}R_{fe\,\,[c}^{\,\,\,\,a}\delta^b_{d]}h^{fe}\Big)$,
$R^{[a}_{\,\,e}\delta^{b]}_{[c}h^e_{\,\,d]}
  - R^{ab}_{\,\,\,\,e[c}h^e_{\,\,d]} + 
R_{fe\,\,\,[c}^{\,\,\,\,[a}\delta^{b]}_{d]}
  h^{fe} = 0$,
with $R_{abcd}$ being the spacetime
Riemann tensor. From these conditions it follows both
that the principal
tensor commutes with the Ricci tensor and that~$\xi_c$ is a Killing
vector field. The principal tensor  having these 
properties is given by
\begin{align}
    &\bm{h} = - (r dr + a^2 \cos\theta\sin\theta d\theta)\wedge dt
    \notag\\&+ a \sin\theta\left(r \sin\theta dr
    + (r^2 + a^2)\cos\theta d\theta \right)\wedge d\phi
    \,.\label{eq:hKerrAds}
\end{align}
The components $h_{ab}$ of the principal tensor
can be extracted directly from Eq.~(\ref{eq:hKerrAds}).
From
$\xi_{b} = \frac{1}{3}\nabla^c h_{c b}$
one then finds
$\xi^a \partial_a = \partial_t$.
The eigenvalues of $h^a_{\,\,b}$ given by
$\pm i x_A$ as referred to above have the following
functions $x_1 = i r$ and $x_2 = a \cos \theta$.



\subsection{Proca field and FKKS Ansatz}

\subsubsection{Kerr-AdS}

The Kerr-AdS spacetime is axisymmetric which by itself is not enough
to separate the equations for the Proca field.  For instance, for the
massless vector field in the Kerr spacetime one needs to use a null
frame to have a neat separation of the
equations~\cite{Teukolsky:1972}.  Extending this result to the massive
case has been a hard task.
Nevertheless, an Ansatz, called the FKKS
Ansatz~\cite{Frolov:2018}, has been discovered and it is able to
separate the Proca equations in Kerr-AdS.  This approach uses the
principal tensor, i.e., a nondegenerate closed conformal Killing-Yano
antisymmetric two tensor.
For Kerr-AdS the Ansatz is given by
\begin{align}
  &A^a = B^{ab}\nabla_b Z\,,
  \label{eq:A1B1}
\end{align}
with $B^{ab}$ being given implicitly in terms of
the metric $g_{bc}$ and the principal tensor $h_{bc}$
by
\begin{align}
  B^{ab}(g_{bc} - \beta h_{bc})
  = \delta^{a}_{c}\,,\label{eq:A1B1xx}
\end{align}
where~$\beta$ is a complex constant
with discrete values to be found,
and~$Z$ is a function
given by
\begin{align}
  Z = R (r) S(\theta) 
  \exp\left(-i\omega t + i m \phi \right)\,.
  \label{ansatzZ}
\end{align}

The polarization tensor defined in
Eq.~(\ref{eq:A1B1xx}) in Kerr-AdS, 
where the metric $g_{ab}$
is taken from the line
element given in Eq.~(\ref{eq:KerrAdSmetric}) 
and the principal tensor  $h_{ab}$
is taken from Eq.~(\ref{eq:hKerrAds}),
can be written
as a symmetric part 
$\bm{B}_{\text{S}}$ and an antisymmetric part
$\bm{B}_{\text{A}}$, namely, 
\begin{align}
  &\bm{B}_{\text{S}} = \frac{\Delta_\Lambda}{q_r \Sigma} \p_r^2
  + \frac{q_\Lambda}{q_\theta \Sigma}\p_\theta^2
  - \frac{1}{q_r \Delta_\Lambda \Sigma}\Big[(r^2 + a^2)\p_t
    + a \p_\phi \Big]^2\nonumber\\
  &+ \frac{1}{\Sigma q_\theta q_\Lambda \sin^2\theta }
  \Big[a \sin^2 \theta \p_t + \p_\phi \Big]^2\,,
  \label{eq:BtensorKerrAdS}
\end{align}
\begin{align}
&\bm{B}_{\text{A}} = \frac{\beta r}{q_r\Sigma}\Big[(r^2 + a^2)
    ( \p_r\p_t-\p_t\p_r ) + a ( \p_r \p_\phi - \p_\phi \p_r )\Big]
  \nonumber\\
  &- \frac{\beta a \sin2\theta}{2\Sigma q_\theta}
  \Big[ a (\p_t \p_\theta - \p_\theta \p_t)
    + \frac{1}{\sin^2\theta}(\p_\phi \p_\theta
    - \p_\theta \p_\phi)\Big]\,,\label{eq:BtensorKerrAdSanti}
\end{align}
with $q_\Lambda = 1 - \frac{a^2}{l^2}\cos^2\theta$,
$q_r = 1 - \beta^2 r^2$, and
$q_\theta = 1 + \beta^2 a^2 \cos^2 \theta$.
The components
${\bm{B}_{\text{S}}}_{ab}$ and ${\bm{B}_{\text{A}}}_{ab}$
can be extracted directly from Eqs.~(\ref{eq:BtensorKerrAdS})
and~(\ref{eq:BtensorKerrAdSanti}), respectively.

One can now put the $B_{ab}$ found in
Eqs.~(\ref{eq:BtensorKerrAdS})
and~(\ref{eq:BtensorKerrAdSanti})
and the Ansatz Eq.~(\ref{ansatzZ})
for $Z$ into the Proca field equation given in 
Eq.~(\ref{eq:A1B1}),
and then into the Proca equations given in 
Eq.~\eqref{eq:Procaeqvacuum}, to get 
the following equations that $R(r)$
and $S(\theta)$ of $Z$ must obey,
\begin{align}
  \p_r\Bigg[\frac{\Delta_\Lambda}{q_r}\p_r R(r) \Bigg]
  + \Bigg[\frac{K_r^2}{q_r \Delta_\Lambda}
  + i \frac{2-q_r}{q_r^2 \beta}\sigma
  + \frac{\mu^2}{\beta^2}\Bigg]R(r) = 0\,,
\label{rofr1}
\end{align}
\begin{align}
  &\frac{1}{\sin\theta}\p_\theta \Bigg[ 
\frac{q_\Lambda\sin\theta}{q_\theta}
  \p_\theta S(\theta)\Bigg]-\nonumber\\
  &\Bigg[\frac{K_\theta^2}{q_\theta q_\Lambda\sin^2\theta }
  + i \frac{2 - q_\theta}{q_\theta^2\beta}\sigma
  + \frac{\mu^2}{\beta^2}\Bigg]S(\theta) = 0\,,
\label{softheta1}
\end{align}
with $K_r = a m - (a^2 + r^2)\omega$,
$K_\theta = m - a \omega \sin^2\theta$, and
$\sigma = a \beta^2 (m - \omega a) - \omega$.
Then, with the solution found, one 
obtains the Proca field as
\begin{align}
A^a(\beta)
=  B^{ab}\partial_b Z(\beta)\,
\label{aaagain}
\end{align}
for each value $\beta$.
Note that 
one may interpret the
different values of $\beta$ found from the
equations as corresponding to
different polarizations. This can be seen by
putting Eq.~(\ref{aaagain}) in the
form
$A^a
= \sum_\beta c_\beta B^{ab}\partial_b Z(\beta)$, where $c_\beta$ are
constants and each $Z(\beta)$
is a different independent solution since they obey different
equations. It is unclear whether or not all solutions can be
described using this Ansatz. It will be seen now that
at least in the Schwarzschild limit there are solutions that are
not described by the Ansatz.

\subsubsection{Schwarzschild-AdS limit}

The special case of Schwarzschild-AdS can be obtained by taking
the limit~$a=0$. The Kerr-AdS line element given in 
Eq.~(\ref{eq:KerrAdSmetric}) reduces to 
the Schwarzschild-AdS line element 
given in 
Eq.~(\ref{eq:fSchwADS}).
In addition, from Eqs.~\eqref{eq:BtensorKerrAdS}
and~\eqref{eq:BtensorKerrAdSanti}, the tensor~$B^{ab}$ in the
nonrotating $a=0$ limit becomes
\begin{align}
  &\bm{B}_{\text{S}} = - \frac{1}{q_r f} \p_t^2 + \frac{f}{q_r}\p_r^2
  + \frac{1}{r^2}\p_\theta^2 + \frac{1}{r^2 \sin^2\theta} \p_\phi^2\,,
\label{eq:symmmode}
\end{align}
\begin{align}
&\bm{B}_{\text{A}} = \frac{\beta r}{q_r} (\p_r \p_t - \p_t \p_r)\,.
\label{eq:antisymmmode}
\end{align}
Equation~\eqref{rofr1}
for~$R(r)$ is now given by
\begin{align}
  \p_r\Big[\frac{r^2 f}{q_r}
  \p_r R\Big] + \Big[\frac{\omega^2 r^2}{f q_r}
  - i \omega \frac{2 - q_r}{q_r^2\beta} + \frac{\mu^2}{\beta^2}
  \Big]R
  =0\,.
  \label{eq:SchwADSRProca}
\end{align}
The angular
equation~\eqref{softheta1} turns into
\begin{align}
  \frac{1}{\sin\theta}\p_\theta\Big[\sin\theta \p_\theta S\Big]
  - \frac{m^2}{\sin^2\theta}S + \Big[i\frac{\omega}{\beta}
    - \frac{\mu^2}{\beta^2}\Big]S=0\,.
    \label{eq:SchwADSSProca2}
\end{align}
The solutions for this equation are the spherical
harmonics~$Y^{\ell m}$.
Thus, it is possible to obtain the expression for the covariant
components of the massive vector field as a function of the
scalar~$R(r)$ and the spherical harmonics~$Y^{\ell m}$ for
each $\beta$.
In the Schwarzschild-AdS background, i.e., $a=0$, Eq.~\eqref{aaagain},
with the help of 
Eqs.~\eqref{eq:symmmode}
and~\eqref{eq:antisymmmode}, give
\begin{align}
  A_a = \Big(-\frac{i \omega}{q_r} +
  \frac{\beta r f}{q_r}\p_r ,\frac{1}{q_r}
  \p_r - i\frac{\omega\beta r}{q_r f} , \p_\theta , \p_\phi \Big)
  R(r) Y^{\ell m}\,,\label{eq:AintermsofR}
\end{align}
where we have dropped the explicit dependence on $\beta$
to not overcrowd the notation. 
Moreover, from 
Eq.~\eqref{eq:SchwADSSProca2}
the values for the parameter~$\beta$
can then be
found by setting
$i\frac{\omega}{\beta} - \frac{\mu^2}{\beta^2} = \ell(\ell + 1)$.
Thus, there are two different values for~$\beta$ for each~$\ell> 0$.
Calling these values 
$\beta_+$ and~$\beta_-$, one has
\begin{align}
    \beta_{\pm} = i \omega \frac{1 \pm \sqrt{1 + \frac{
    4 \mu^2 \ell(\ell + 1)}{\omega^2}}}{2 \ell (\ell + 1)}\,.
    \label{eq:betapm}
\end{align}
The two different~$\beta$ values each correspond to a different
polarization.
A further  analysis on the expression suggests
that~$\beta_-$ describes the scalar-type polarization, since
setting the Proca mass~$\mu = 0$
makes it vanish as one expects for the massless Maxwell field. 
Moreover, setting the Proca~$\ell$ to zero
yields a definite value for $\beta_-$, namely 
$\beta_{\text{monopole}} = - i \frac{\mu^2}{\omega}$,
so that $\beta_-$ is the polarization that contains
the monopole case.
The case $\beta_+$ describes then the vector-type 
polarization, since
setting the Proca mass~$\mu = 0$
yields 
$\beta_+ = i \omega \frac{1}{ \ell (\ell + 1)} $,
as one expects for the massless Maxwell field,
whereas setting the Proca~$\ell$ to zero, $\ell=0$,
gives an infinite $\beta$ which seems
to have no meaning.

Two important features can be drawn from the FKKS Ansatz of
Eq.~\eqref{eq:AintermsofR}.  
The first is that there is a natural
decoupling of the two polarizations related to the electric modes, in
contrast to the VSH method of
Sec.~\ref{sec:ProcafieldSchwADS}. 
The second feature is that when comparing
Eq.~\eqref{eq:AintermsofR} with Eq.~\eqref{eq:SchwParam}, it can be
seen that the Ansatz in the Schwarzschild-AdS limit, $a= 0$, does not
describe the vector-type polarization related to the magnetic modes,
i.e., it does not describe the function $u_{(3)}$ in
\eqref{eq:SchwParam}, again in
contrast to the VSH method  of
Sec.~\ref{sec:ProcafieldSchwADS}.  By inspecting the
principal tensor in
Eq.~\eqref{eq:hKerrAds}, two of the eigenvalues of~$h^a_{\,\,b}$
are given by~$\pm i x_2 = \pm i a \cos{\theta}$. By taking the
Schwarzschild-AdS limit, both eigenvalues go to zero and so the
principal tensor becomes degenerate. This violates the initial
requirement that the principal tensor needs to be nondegenerate in
order to characterize all the symmetries of the spacetime. This fact
surely has implications on the absence of the magnetic modes of the
Proca field in the FKKS approach.

\subsection{Quasinormal modes in Schwarzschild-AdS in 
the FKKS method}
\label{sec:quasinormal2}

\begin{table*}[t]
\centering

        \begin{subtable}{1\textwidth}
    \centering
    \begin{tabular}{|c|c|c|}
        \hline
                        & $r_+=l$ & $r_+ = 100 l$\\\hline
        $\mu l$ & $\omega l$ (FKKS) & $\omega l$ (FKKS)\\ \hline
        $0.01$ & $3.330 - 2.489\,\,i$ & $184.968 - 266.395\,\,i$  \\ \hline
        $0.10$ & $3.339 - 2.501\,\,i$ &   $185.578 - 267.524 \,\,i$ \\ \hline
        $0.20$ & $3.362 - 2.534\,\,i$ &  $187.355 - 270.817\,\,i$ \\ \hline
        $0.40$ & $3.444 - 2.652\,\,i$ & $193.650 - 282.498\,\,i$ \\ \hline
        $0.50$ & $3.498 - 2.729\,\,i$ & $197.761 - 290.138\,\,i$ \\ \hline
        \end{tabular}
        \caption{Scalar-type polarization ($\beta_-$).\label{tab:QNMmBmSAdS}}
\end{subtable}

\vskip 0.75cm

\begin{subtable}{1\textwidth}
\centering
    \begin{tabular}{|c|c|c|}
        \hline
                        & $r_+=l$ & $r_+ = 100 l$\\\hline
        $\mu l$ &   $\omega l$ (FKKS) & $\omega l$ (FKKS)\\ \hline
        $0.01$ & $1.554 - 0.542 \,\,i$ &  $0 - 149.984\,\, i$\\ \hline
        $0.10$ & $1.557 - 0.552 \,\,i$ & $0 - 152.099\,\,i$  \\ \hline
        $0.20$ & $1.568 - 0.584\,\,i$ & $0 - 158.432\,\,i$ \\ \hline
        $0.30$ & $1.584 - 0.633\,\,i$ & $0 - 168.817\,\,i$ \\ \hline
        $0.40$ & $1.606 - 0.699 \,\,i$ & $0 - 183.291\,\,i$ \\ \hline
        $0.50$ & $1.632 - 0.777\,\, i$ & $0 - 202.684\,\,i$ \\ \hline
        \end{tabular}
         \caption{Vector-type polarization ($\beta_+$).\label{tab:QNMmBpSAdS}}
        \end{subtable}\vspace{5mm}
\caption{
Quasinormal mode frequencies $\omega l$ of the Proca field electric
modes with~$\ell = 1$, using the FKKS method in Schwarzschild-AdS
for~$r_+ = l$ and $r_+ = 100l$ and for several values of the Proca
field mass~$\mu l$. (a) The frequencies of the scalar-type polarization
$\beta_-$ of the electric modes are displayed.  (b) The frequencies of
the vector-type polarization $\beta_+$ of the electric modes are
displayed.
\label{tab:QNMFKKS}}
\end{table*}

The aim now is to solve Eq.~\eqref{eq:SchwADSRProca}
for $R$.
Note beforehand that, as proved 
analytically in \cite{thesis},
the Ansatz of Eq.~\eqref{eq:AintermsofR}
continues to obey the Proca equations given in
Eqs.~\eqref{eq:SchwEM2better} and \eqref{eq:SchwEM3}, and the Bianchi
identity given in Eq.~\eqref{eq:LorenzConditionSchw}, by considering
the correspondence $u_{(0)} = - \frac{i\omega r}{q_r} R(r) +
\frac{\beta r^2 f}{q_r} \p_r R(r)$, $u_{(1)} = \frac{r f}{q_r} \p_r
R(r) - i \frac{\omega \beta r^2}{q_r} R(r)$, and $u_{(2)} = \ell(\ell
+ 1) R(r)$, for each $\beta$.  This remarkable correspondence between
the VSH method and the FKKS method given by these transformations,
decouples the two polarizations in the Proca equations with the
relevant scalar function still obeying a second-order partial
differential equation, a result
that has also been shown in \cite{Percival:2020skc}.

We can now solve Eq.~\eqref{eq:SchwADSRProca}
for $R$.
Indeed, in Eq.~\eqref{eq:SchwADSRProca},
we must swap the derivatives in $r$ in terms of derivatives in
$r_*$. Afterwards by multiplying 
the obtained equation 
by a factor $\frac{q_r f}{r^2}$, we
obtain Eq.~\eqref{eq:SchwEM3} with $u_{(1)}$ and $u_{(2)}$
given by the correspondence above, i.e., 
$u_{(1)} = \frac{r f}{q_r} \p_r R(r)
  - i \frac{\omega \beta r^2}{q_r} R(r)$,
and $u_{(2)} = \ell(\ell + 1) R(r)$.
We then
write $R= \mathcal{R}
e^{-i\omega(t + r_*)}$ and
substitute it into  Eq.~\eqref{eq:SchwEM3} with
the given correspondence to yield
\begin{align}
  \Bigg[(x - x_+) v(x)\p_x^2  + w(x)\p_x
    + \frac{y(x)}{x-x_+} \Bigg]\mathcal{R} = 0\,,
    \label{eq:HorowitzeqR}
\end{align}
where again $x=\frac{1}{r}$ and $x_+ = \frac{1}{r_+}$,
and the polynomials are
\begin{align}
  &v(x) = x^2 \frac{f}{r^2} \frac{(x^2 - \beta^2)}{x-x_+}\,,
  \label{eq:stuFKKS0}\\
&w(x) = (x^2 - \beta^2) (2 i \omega x^2 + 2 x^3 - 6 M x^4)
  - 2 x^3 \frac{f}{r^2}\,,
\label{eq:stuFKKS1}\\
&y(x) = (x - x_+)\Big[(x^2 - \beta^2)(\mu^2 + \ell(\ell + 1)x^2)
    \notag\nonumber\\
  &\quad- 2 i \omega (x^3 + \beta x^2)\Big]\,.
  \label{eq:stuFKKS}
\end{align}
The clear differences between the above polynomials and the ones in
Eqs.~\eqref{eq:poly}-\eqref{eq:poly2} are, first, the dependence on the
parameter $\beta$, which is characteristic for each polarization
and, second, the higher order in $x$. The explicit dependence of the
polynomials on $\omega$, without considering the $\beta$ dependence,
at first glance will be at most linear for the same reasons stated for
the polynomials in Eqs.~\eqref{eq:poly}-\eqref{eq:poly2}.
Another difference, 
perhaps more concealed, 
is that when introducing $\beta$,
using Eq.~\eqref{eq:betapm}, into 
Eq.~\eqref{eq:HorowitzeqR}
together with Eqs.~\eqref{eq:stuFKKS0}-\eqref{eq:stuFKKS}, 
one finds that the polynomials will also
have a different dependence on $\omega$, $\ell$, and $\mu$ compared
with the ones in Eqs.~\eqref{eq:poly}-\eqref{eq:poly2}. The polynomials
defined in Eqs.~\eqref{eq:stuFKKS0}-\eqref{eq:stuFKKS} can be expanded
around~$x_+$ so that it is possible
to write
\begin{align}
&v(x) = \sum_{j=0}^{\infty} v_j
(x-x_+)^j\,,
\label{v}\\
&w(x) = \sum_{j=0}^{\infty} w_j
(x-x_+)^j\,,
\label{w}\\
&y(x) = \sum_{j=0}^{\infty} y_j
(x-x_+)^j\,,
\label{y}
\end{align}
respectively, where $v_j$, $w_j$, and $y_j$ are expansion coefficients
that vanish for $j$ higher than some value. Then the scalar
$\mathcal{R}$ can be expanded as
\begin{equation}
\mathcal{R} = \sum_{n=0}^\infty a_{(R)n}
(x-x_+)^n\,,
\label{R}
\end{equation}
where the coefficients $a_{(R)}$
are calculated using the recurrence
relation
\begin{align}
  &a_{(R)n} = - \frac{1}{P_n} \sum_{j=0}^{n-1}
  \Big(j(j-1) v_{n-j} + j w_{n-j} + y_{n-j} \Big)a_{(R)j}\,,
  \label{eq:recurrencesimpleR}
\end{align}
where $P_n$ is given by 
\begin{align}
P_n =  n(n-1) v_0 + n w_0\,,
\end{align}
and the
$v_{j}$, $w_{j}$, and $y_{j}$ are the coefficients of the expansion in
Eqs.~\eqref{v}-\eqref{y}, considering the polynomials in
Eqs.~\eqref{eq:stuFKKS0}-\eqref{eq:stuFKKS}.
In putting the problem in this way 
we have just shown that 
the quasinormal modes derived from 
the FKKS Ansatz in the nonrotating limit 
of Kerr-AdS, i.e., for Schwarzschild-AdS, 
can now be computed by
applying the Horowitz-Hubeny numerical procedure.
This means that
the quasinormal modes can be computed by
requiring that $R$ vanishes at $x\rightarrow 0$, thus
\begin{align}
\sum_{j=0}^N a_{(R)j} (-x_+)^j = 0\,,
\end{align}
where $N$ is formally infinite but for numerical purposes is a large
integer. 
We do not present the mode frequencies for $\ell = 0$ which would be
taken from $\beta_-$, but they were calculated by us using this method
and they agree with frequencies calculated using the VSH method of
Sec.~\ref{sec:ProcafieldSchwADS} and with the results
in~\cite{Konoplya:2006}.  The numerical calculations of the
quasinormal modes for $\ell = 1$ for Schwarzschild-AdS in the FKKS
method are presented in Table~\ref{tab:QNMFKKS}.  These quasinormal
modes were computed for each of the two values of $\beta$ taken from
Eq.~\eqref{eq:betapm}.  Since $\beta_-$ can be identified as
corresponding to the scalar-type polarization, and $\beta_+$ can be
identified as corresponding to the vector-type polarization, the modes
for each polarization are easily distinguished.  The mode frequencies
for the scalar-type polarization $\beta_-$ are shown in
Table~\ref{tab:QNMmBmSAdS} and the mode frequencies for vector-type
polarization $\beta_+$ are shown in Table~\ref{tab:QNMmBpSAdS}.  It
was taken $N=40$, and the method seems to converge even though the
polynomials given in Eqs.~\eqref{eq:stuFKKS0}-\eqref{eq:stuFKKS} have
higher order dependence in $x$.  The values of the quasinormal mode
frequencies for other higher monotones, higher values of $\mu$ and
$r_+$, are found in \cite{thesis}.  We must emphasize that the
quasinormal modes for the other vector-type polarization, encoded in
$u_{(3)}$, cannot be found since the FKKS Ansatz in the
Schwarzschild-AdS limit does not describe the magnetic modes.


\section{Comparison of results between the VSH method and
the FKKS method\label{sec:compare}}



\begin{table*}[t]
\centering
        \begin{subtable}{1\textwidth}
    \centering
    \begin{tabular}{|c|c|c|c|c|}
        \hline
                        & \multicolumn{2}{c|}{$r_+=l$} & \multicolumn{2}{c|}{$r_+ = 100 l$}\\\hline
        $\mu l$ &   $\omega l$ (VSH) & $\omega l$ (FKKS) & $\omega l$ (VSH) & $\omega l$ (FKKS)\\ \hline
        $0.01$ & $3.331 -  2.489\,\, i$ & $3.330 - 2.489\,\,i$  &   $184.968 - 266.394\,\, i$ & $184.968 - 266.395\,\,i$  \\ \hline
        $0.10$ & $3.339 - 2.500\,\, i$ & $3.339 - 2.501\,\,i$ &   $185.604 - 267.461\,\, i$ & $185.578 - 267.524 \,\,i$ \\ \hline
        $0.20$ & $3.362 -  2.531\,\, i$ & $3.362 - 2.534\,\,i$ &    $187.452 - 270.612\,\, i$ & $187.355 - 270.817\,\,i$ \\ \hline
        $0.40$ & $3.446 -  2.645\,\, i$ & $3.444 - 2.652\,\,i$ &  $193.925 - 282.119\,\, i$ & $193.650 - 282.498\,\,i$ \\ \hline
        $0.50$ & $3.501 - 2.722\,\, i$ & $3.498 - 2.729\,\,i$ & $198.077 - 289.799\,\, i$ & $197.761 - 290.138\,\,i$ \\ \hline
        \end{tabular}
        \caption{Scalar-type polarization.\label{tab:QNMmBmSAdScomparison}}
\end{subtable}

\vskip 0.75cm

\begin{subtable}{1\textwidth}
\centering
    \begin{tabular}{|c|c|c|c|c|}
        \hline
                        & \multicolumn{2}{c|}{$r_+=l$} & \multicolumn{2}{c|}{$r_+ = 100 l$}\\\hline
        $\mu l$ &   $\omega l$ (VSH) & $\omega l$ (FKKS) & $\omega l$ (VSH) & $\omega l$ (FKKS)\\ \hline
        $0.01$ & $1.554 - 0.542 \,\,i$ & $1.554 - 0.542 \,\,i$ &   $0 - 149.984\,\, i$ &  $0 - 149.984\,\, i$\\ \hline
        $0.10$ & $1.557 - 0.552\,\, i$ & $1.557 - 0.552 \,\,i$ &   $0 - 152.099\,\, i$ & $0 - 152.099\,\,i$  \\ \hline
        $0.20$ & $1.568 -  0.583\,\, i$ & $1.568 - 0.584\,\,i$ &    $0 - 158.432\,\, i$ & $0 - 158.432\,\,i$ \\ \hline
        $0.30$ & $1.585 -  0.633\,\, i$ & $1.584 - 0.633\,\,i$ &  $0 - 168.817\,\, i$ & $0 - 168.817\,\,i$ \\ \hline
        $0.40$ & $1.607 -  0.699\,\, i$ & $1.606 - 0.699 \,\,i$ &  $0 - 183.291\,\, i$ & $0 - 183.291\,\,i$ \\ \hline
        $0.50$ & $1.634 - 0.777\,\, i$ & $1.632 - 0.777\,\, i$ & $0 - 202.684\,\, i$ & $0 - 202.684\,\,i$ \\ \hline
        \end{tabular}
         \caption{Vector-type polarization.\label{tab:QNMmApSAdScomparison}}
        \end{subtable}\vspace{5mm}

\caption{
Quasinormal mode frequencies $\omega l$ of the Proca field electric
modes $u_{(0)}$, $u_{(1)}$, and $u_{(2)}$ with~$\ell = 1$,
comparing the VSH and FKKS methods in Schwarzschild-AdS for~$r_+ =
l$ and $r_+ = 100l$ and for several values of the Proca field
mass~$\mu l$. (a) The frequencies of the scalar-type polarization 
of the electric modes in both methods are displayed. 
(b) The frequencies of the vector-type polarization 
of the electric modes in both methods are displayed.
The numbers are taken from Tables~\ref{table1u1u2u3} and
\ref{tab:QNMFKKS}.
   \label{tab:QNM}}
\label{tab:QNMmmSAdScomparison}
\end{table*}



The computations of the quasinormal modes were 
performed numerically using
\textit{Mathematica}.
In the computations we have put $N = 40$, and higher $N$ 
would not change the results as presented.

The quasinormal modes for the monopole, $\ell = 0$, using
both the VSH
method and the FKKS
method were calculated by us and are consistent
with~\cite{Konoplya:2006}, so we do not need
to present them here.

The quasinormal modes for  $\ell = 1$, 
in the electric mode sector
using
both the VSH
and the FKKS
Ans\"{a}tze are put together 
and displayed in 
Table~\ref{tab:QNMmmSAdScomparison}
for a comparison between both methods.
The values shown 
in Table~\ref{tab:QNMmmSAdScomparison}
are taken directly from Tables~\ref{table1u1u2u3} and
\ref{tab:QNMFKKS}.
In  Table~\ref{tab:QNMmBmSAdScomparison}
the 
quasinormal frequencies 
for  the scalar-type polarization
are shown, and 
in Table~\ref{tab:QNMmApSAdScomparison}
the 
quasinormal frequencies 
for the vector-type polarization are shown.
The magnetic sector does not appear in 
the FKKS method so there is no possibility
of comparison in this sector.

In the VSH method it is hard to
distinguish the polarizations in the electric modes for low $r_+$,
an example being $r_+=l$. 
On the other hand, for large $r_+$,
$r_+ \geq 100l$, one can distinguish
them  by comparison
with the
frequencies of the modes in the magnetic sector, 
since here both vector-type
polarizations have modes with negligible real frequency.
In the FKKS method polarization is well characterized by the
different values of~$\beta$, allowing a direct
distinction. By
convenience, our strategy to distinguish the electric modes in the VSH
method was to use the frequencies given by
the FKKS Ansatz and check if
they were present.

The massless limit $\mu=0$ of the quasinormal modes shall now be
analyzed in detail. As referred to in
Sec.~\ref{sec:ProcafieldSchwADS}, when analyzing the quasinormal modes
in the VSH method, the $\mu=0$ scalar-type polarization becomes
nonphysical but its quasinormal modes, which correspond to the
ones of massless Klein-Gordon scalar field as it can be checked both
analytically and numerically, will not disappear automatically. We now
proceed to explain how this polarization can be removed. We notice
that for $\mu=0$, i.e., for a Maxwell field, there are only two
physical degrees of freedom, and the corresponding equations are
governed by the $\mu=0$ Proca equations with the Bianchi identity of
Eq.~\eqref{eq:LorenzConditionSchw} now being a gauge condition,
specifically, the Lorenz condition.  But even with the Lorenz
condition being imposed, one is still left with a spurious degree of
freedom which corresponds to the contribution of a gradient of a
scalar field which obeys the massless Klein-Gordon equation.  By
counteradding the gradient of this scalar field with the same
modes, one is able to remove the spurious degree of freedom and reset
the results for the quasinormal modes of a Maxwell field in
Schwarzschild-AdS.  As referred to in Sec.~\ref{sec:FKKS}, when
analyzing the quasinormal modes in the FKKS method, the $\mu=0$
scalar-type polarization becomes nonphysical as it should.  Here, we
note that Eq.~\eqref{theequation1} is equivalent to
Eq.~\eqref{eq:SchwEM3} with the correspondence $u_{(1)} = \frac{r
f}{q_r} \p_r R(r) - i \frac{\omega \beta r^2}{q_r} R(r)$, $u_{(2)} =
\ell(\ell + 1) R(r)$.  Setting $\mu = 0$ one has from
Eq.~\eqref{eq:betapm} that $\beta_-=0$, and then Eq.~\eqref{eq:betapm}
turns into an equation for a massless Klein-Gordon scalar field.  The
same reasoning that we have performed for the VSH method applies now,
and again one is able to remove the spurious degree of freedom and
reset the results for the quasinormal modes of a Maxwell field in
Schwarzschild-AdS.

The maximum relative deviation  between the quasinormal
frequencies of both treatments
was found to be $0.2\%$
which is the case of the last row of
Tab.~\ref{tab:QNMmBmSAdScomparison}
for $r_+=l$. The relative deviation $\sigma$
was calculated
through the formula $\sigma(\%) = \sqrt{\frac{\abs{
{{\omega_{}}_{}}_{\rm VSH} -
{{\omega_{}}_{}}_{\rm FKKS}}
}{\abs{
{{\omega_{}}_{}}_{\rm VSH} 
}}}\times100$. 
This confirms that the electric 
modes of the VSH method
are described in 
the FKKS Ansatz.
We should also add that the 
Horowitz-Hubeny numerical procedure 
applied to the VSH and the FKKS methods for 
the electric
modes in obtaining the quasinormal frequencies
work fine
with the
FKKS method converging faster see the Appendix~\ref{appa}.
Furthermore, the VSH method does not capture part of the
quasinormal modes for the same $N$ of the FKKS method. These modes
certainly should appear for
the ideal $N \rightarrow \infty$ limit.
Surely, this adds value to the
FKKS decomposition of the vector $A_a$, Eq.~\eqref{eq:AintermsofR}, in
the electric mode sector to decouple the two polarizations.

In \cite{thesis}, the quasinormal modes for each method were computed
for higher monotones for different values of $\mu l$ and for
different values of $\frac{r_+}{l}$. There were higher monotones of the
quasinormal modes of~$u_{(0)}$,~$u_{(1)}$, and~$u_{(2)}$ in the VSH
method that could not be found, but which were present within the
FKKS Ansatz. The same occurred when computing the frequencies for
higher values of $\mu$, for example $\mu l = 8$. The convergence of
the VSH method, generalized here for a system of equations,
was not
demonstrated, thus it may be possible that the fact of having two
polarizations in the system requires a much higher value of~$N$ so
that these monotones can be found in the VSH method. To show
rigorously that both the VSH and FKKS methods describe the same
quasinormal modes, we would need to compare every frequency, for all
the space of variables $\mu l$ and $\frac{r_+}l$, and also compare all
the higher monotones. This is an impossible task but the fact that it
was shown that $R(r)$ obeys Eqs.~\eqref{eq:SchwEM2better},
\eqref{eq:SchwEM3}, and \eqref{eq:LorenzConditionSchw} with the
correspondence $u_{(0)} = - \frac{i\omega r}{q_r} R(r) + \frac{\beta
r^2 f}{q_r} \p_r R(r)$, $u_{(1)} = \frac{r f}{q_r} \p_r R(r) - i
\frac{\omega \beta r^2}{q_r} R(r)$, and $u_{(2)} = \ell(\ell + 1)
R(r)$ substantiates that such is the case \cite{thesis}.

\section{Conclusions}
\label{conc}

We have separated the Proca equations in a Schwarzschild-AdS spacetime
by using the VSH method, which employs vector spherical harmonics in a
spherically symmetric spacetime, in our case in the Schwarzschild-AdS
spacetime. Specifically, the Proca field was taken to 
satisfy an Ansatz in terms
of the vector spherical harmonics and of 
time and radial dependent functions $u_{(i)}$, with
$i=0,1,2,3$. These functions can be
classified into electric modes ($u_{(0)}$, $u_{(1)}$,
and $u_{(2)}$)
and magnetic modes ($u_{(3)}$).  The Proca equations thus separate and
give a system of partial differential equations which are coupled
for the electric modes and decoupled for the magnetic modes. The
dynamical solutions of the system will have three degrees of freedom
which we call polarizations, and each polarization will have its sets
of eigenvectors. The polarization is of scalar-type if the set of
eigenvectors corresponding to it behave similarly to a scalar,
otherwise the polarization is vector-type. The electric modes have a
scalar and a vector-type polarization, whereas the
magnetic modes have the remaining vector-type polarization. Since the
equations for the electric modes are coupled, it is hard to
distinguish their polarizations. In the massless limit, the equations
can be decoupled without increasing their order and the two
vector-type polarizations degenerate, while the scalar-type 
polarization is described by a gradient of a massless 
scalar field which 
obeys the Klein-Gordon equation.  The quasinormal modes for the
Proca field can then be found and were displayed.  In the limit of
massless field with $r_+=100 l$, the quasinormal modes of the
vector-type polarizations approach the values given in
\cite{Cardoso:2001}.  For the monopole mode $\ell=0$ we have not shown
the values but our calculations agree with those in
\cite{Konoplya:2006}.  The quasinormal modes were found using an
extension of the Horowitz-Hubeny numerical procedure
\cite{Horowitz:2000} for the electric modes and the original method to
the magnetic modes.

We have also studied the Proca equations in a Schwarzschild-AdS
spacetime using the FKKS method.  The complete set of the spacetime
symmetries is generated by the principal tensor and
allows a separation
of the Proca equations for generic spinning geometries, in particular
in the Kerr-AdS spacetime \cite{Frolov:2018}.  The Ansatz describes
the Proca field as a contraction of the polarization tensor with the
gradient of a complex scalar and enables the Proca equations to reduce
to an angular and radial equation for that complex scalar.  The
polarization tensor depends on the principal tensor and a complex
constant $\beta$, whose discrete values are determined by the
equations and each value corresponds to a different polarization.  It
remains unclear whether the FKKS Ansatz captures all the degrees of
freedom of the Proca vector field. In order to study this issue, an
analysis of the Proca system in the nonrotating limit of the Kerr-AdS
spacetime, i.e., in the Schwarzschild-AdS spacetime, was made.  We
analyzed this Ansatz in Schwarzschild-AdS to check if it is able to
describe all the polarizations. We showed that the FKKS Ansatz in the
Schwarzschild-AdS limit describes two polarizations of the massive
vector field, namely the scalar-type and the vector-type polarizations
corresponding to the electric modes of the field and it was verified
that the method allows for an easier identification of each
polarization.  Moreover, for these electric modes an analytical
correspondence between the VSH method and the FKKS method was
obtained, revealing a remarkable transformation that decouples the two
polarizations in the Proca equations with the relevant scalar function
still obeying a second-order partial differential equation.
In the massless limit, we observed that the complex
scalar field of the FKKS Ansatz associated to the scalar-type
polarization obeys the Klein-Gordon equation and the expression for
the vector field is equivalent to the gradient of this complex
scalar. Thus, the FKKS method in the massless limit yields the same
result as the VSH method.  On the other hand, the FKKS Ansatz does
not capture the magnetic modes.  Since it is known that
magnetic modes are present in the Kerr geometry in the FKKS
Ansatz~\cite{Dolan:2018dqv, Baumann:2019eav,Percival:2020skc},
the
reason for the absence of the magnetic modes may be due to the
degeneracy of the principal tensor in the nonrotating limit.  The
quasinormal modes of the electric sector of the Proca field with the
FKKS Ansatz were computed using the Horowitz-Hubeny numerical
procedure.

We performed a numerical comparison of the quasinormal modes of the
electric sector that were obtained by the VSH method and the FKKS
method.  The quasinormal modes of the electric sector in both methods
coincide well, having a maximum relative deviation of~$0.2\%$. Even
though only the fundamental quasinormal modes are displayed here for
$\ell=1$, this corroborates that the FKKS method is not only able to
describe the electric modes but also is able to decouple both
polarizations naturally in the electric mode
sector. In the massless limit, the quasinormal modes
for the scalar-type polarization do not vanish in both methods,
coinciding with the quasinormal modes of a Klein-Gordon scalar
field. Nevertheless, these modes are nonphysical since they can be
removed by the gauge freedom. Since the FKKS Ansatz does not describe
the magnetic modes, the quasinormal modes associated to this sector
cannot be compared.

Further study of the polarizations described by the FKKS Ansatz in
spinning geometries should be undertaken. For the Kerr metric an
analytical comparison between the Teukolsky method and the FKKS method
in the massless limit was presented in~\cite{Dolan:2018dqv}, and so it
would be interesting to see an extension of such a comparison for
Kerr-AdS, and even for Kerr-dS.  This is achievable since the
Newman-Penrose formalism used by Teukolsky in Kerr can in principle be
extended to Kerr-AdS.

\acknowledgments

This work was supported through the European Research Council
Consolidator Grant No.~647839, the Portuguese 
Science Foundation FCT Project No.~IF/00577/2015,
the FCT Project No.~PTDC/MAT-APL/30043/2017, the FCT
Project~No.~UIDB/00099/2020, and the FCT
Project~No.~UIDP/00099/2020.
This work has received funding from the European Union's Horizon
2020 research and innovation programme under the Marie
Sklodowska-Curie grant agreement No.~101007855.
The authors would like to acknowledge networking support by the
GWverse COST Action CA16104, ``Black holes, gravitational waves and
fundamental physics.''


\appendix

\section{Comparison of numerical
convergence between the VSH and FKKS methods }
\label{appa}

\begin{figure*}[t]
    \centering
    \begin{subfigure}{0.4\textwidth}
    \centering
    \includegraphics[width=\textwidth]{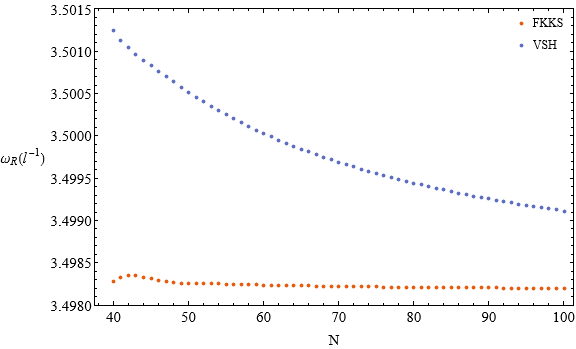}
    \caption{
    The frequency $\omega_R$, the real part of $\omega$, as 
    a function of
    $N$.}
    \label{fig:mu050rp1Re}
    \end{subfigure}
    \begin{subfigure}{0.4\textwidth} 
    \centering
    \includegraphics[width=\textwidth]{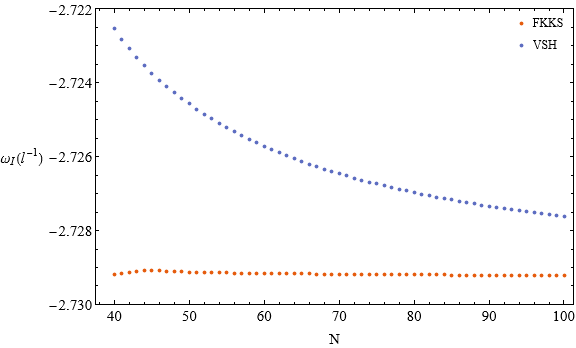}
    \caption{The frequency $\omega_I$, the
    imaginary part of $\omega$, as a function of
    $N$.}
    \label{fig:mu050rp1Im}
    \end{subfigure}
    \caption{Quasinormal mode of the scalar-type polarization of the
    Proca field for $\ell = 1$, $\mu l = 0.50$, $r_+ = l$ computed
    with varying computational number $N$, for the VSH method (blue)
    and the FKKS method (orange). The values in blue seem to converge
    to the values of orange.}
    \label{fig:mu050rp1}
\end{figure*}

\begin{figure*}
    \centering
    \begin{subfigure}{0.4\textwidth}
    \centering
    \includegraphics[width=\textwidth]{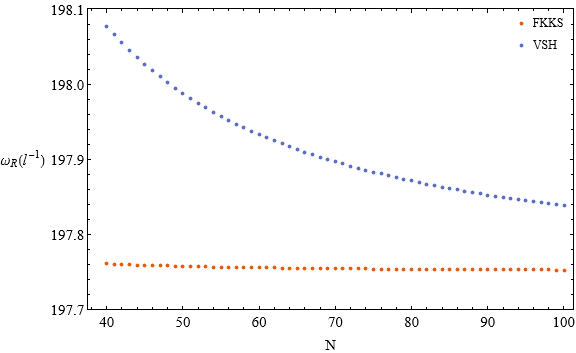}
    \caption{The frequency $\omega_R$, the real part of $\omega$,
    as a function of
    $N$.}
    \label{fig:mu050rp100Re}
    \end{subfigure}
    \begin{subfigure}{0.4\textwidth} 
    \centering
    \includegraphics[width=\textwidth]{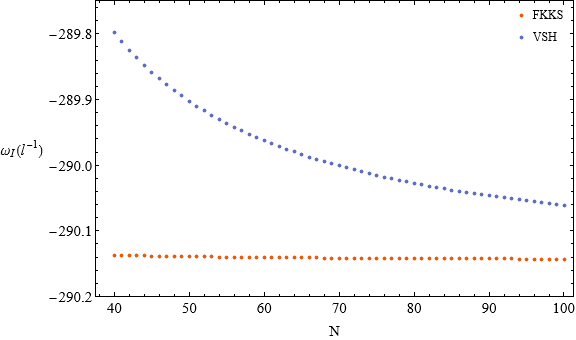}
    \caption{The frequency $\omega_I$, the
    imaginary part of $\omega$, as a function of
    $N$.}
    \label{fig:mu050rp100Im}
    \end{subfigure}
    \caption{Quasinormal mode of the scalar-type polarization of the
    Proca field for $\ell = 1$, $\mu l = 0.50$, $r_+ = 100 l$ computed
    with varying computational number $N$, for the VSH method (blue)
    and the FKKS method (orange). The values in blue seem to converge
    to the values of orange.}
    \label{fig:mu050rp100}
\end{figure*}

In Sec.\ref{sec:compare}, when comparing the VSH and FKKS methods we
made notice that the FKKS method converges faster.
To see this convergence explicitly we 
show two figures, 
Figs.~\ref{fig:mu050rp1} and \ref{fig:mu050rp100},
where the quasinormal modes corresponding to the last row of
Tab.~\ref{tab:QNMmBmSAdScomparison} are computed with varying $N$. 
The
value obtained by the VSH method seems to converge to the value
obtained by the FKKS method. Even though the value corresponding to
the FKKS method also changes, this change only occurs in the
fifth
significant digit. It was not possible to compute for higher $N$ since
it requires higher machine precision.


\end{document}